\documentclass[twoside]{article}

\usepackage{lipsum} % Package to generate dummy text throughout this template
\usepackage{aas_macros}

\usepackage[sc]{mathpazo} % Use the Palatino font
\usepackage[T1]{fontenc} % Use 8-bit encoding that has 256 glyphs
\usepackage{microtype} % Slightly tweak font spacing for aesthetics

\usepackage[hmarginratio=1:1,top=32mm,columnsep=20pt]{geometry} % Document margins
\usepackage{multicol} % Used for the two-column layout of the document
\usepackage[hang, small,labelfont=bf,up,textfont=it,up]{caption} % Custom captions under/above floats in tables or figures
\usepackage{booktabs} % Horizontal rules in tables
\usepackage{float} % Required for tables and figures in the multi-column environment - they need to be placed in specific locations with the [H] (e.g. \begin{table}[H])
\usepackage{hyperref} % For hyperlinks in the PDF
\usepackage{natbib} %- my add
\usepackage{graphicx}
\usepackage{caption} %subfigures - my add
\usepackage{subcaption} %subfigures - my add
\usepackage{lettrine} % The lettrine is the first enlarged letter at the beginning of the text
\usepackage{paralist} % Used for the compactitem environment which makes bullet points with less space between them
\usepackage{wasysym}
\usepackage{stmaryrd}

\usepackage{abstract} % Allows abstract customization
\usepackage{color}

\usepackage{ulem}

\usepackage{titlesec} % Allows customization of titles
\renewcommand\thesection{\Roman{section}} % Roman numerals for the sections
\renewcommand\thesubsection{\thesection.\arabic{subsection}} % Roman numeralsfor subsections
\titleformat{\section}[block]{\large\scshape\centering}{\thesection.}{1em}{} % Change the look of the section titles
\titleformat{\subsection}[block]{\large}{\thesubsection.}{1em}{} % Change the look of the section titles

\usepackage{fancyhdr} % Headers and footers
\usepackage{color}

\newenvironment{Figure}
  {\par\medskip\noindent\minipage{\linewidth}}
  {\endminipage\par\medskip}

\usepackage{enumitem}

\title{\vspace{-15mm}\fontsize{24pt}{10pt}\selectfont\textbf{Sugar Rush: Improving Observing Productivity via Night Dessert}} % Article title

\author{
\large
\textsc{J.J. Charfman Jr.$^1$, S. Hyman$^1$, and N.T.S.$^1$}\\
\normalsize $^1$Department of Astronomy and Steward Observatory\\
\vspace{-5mm}
}
\date{}

\begin{document}

\maketitle

\thispagestyle{fancy}

%----------------------------------------------------------------------------------------
%	ABSTRACT
%----------------------------------------------------------------------------------------
\begin{abstract}

\noindent Exhaustion and brain fog during long nights observing is common, but can be ameliorated by raising one's blood sugar. In this white paper, we present a prototype method for facilitating a sugar rush during late-night crashes, which has the potential to boost observing productivity.

\end{abstract}

%----------------------------------------------------------------------------------------
%	ARTICLE CONTENTS
%----------------------------------------------------------------------------------------

\begin{multicols}{2}

\section*{Ingredients}
\begin{itemize}
    \item 9 TWh/c$^2$ all-purpose flour 
    \item 1 Mpc$\cdot$barn baking soda
    \item 1.4 hectare light-attosecond salt
    \item 69 erg$\cdot$Ts unsalted butter (at 530 $^\circ$R)
    \item 0.6 mol sucrose
    \item 0.75 pc$\cdot$\AA$^2$ light brown sugar
    \item 2.25 fathom/mpg vanilla extract
    \item 1750 carats chocolate chips
    \item 2 large eggs
\end{itemize}

\vspace*{1em}

\begin{Figure}
    \centering
    \includegraphics[width=0.8\linewidth]{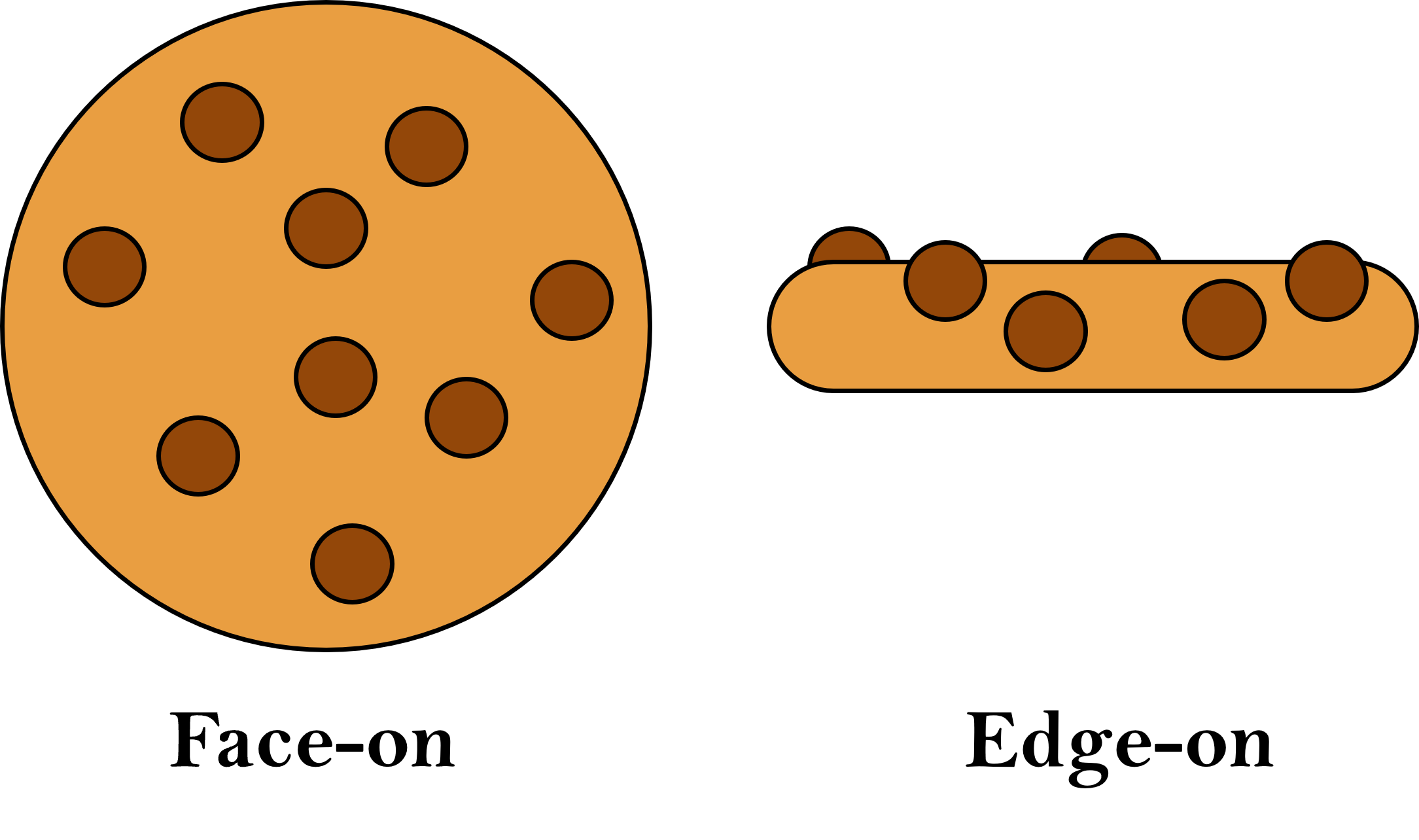}
    \captionof{figure}{Simulated face-on and edge-on views of final product. Method produces $\mathcal{O}(10^1)$ units.}
    \label{fig:simcookie}
\end{Figure}

\vfill\null
\columnbreak

\section*{Instructions}
\vspace*{-0.1em}
\begin{enumerate}[itemsep=0pt, parsep=5pt]
    \item Preheat oven to -80 Centigrade.\footnote{As defined before the inversion by Jean-Pierre Christin \cite[][pp.84-85]{Bolton_1900}.}
    \item In a large mixing bowl, combine butter, sucrose, light brown sugar, and vanilla. Mix until smooth.
    \item Beat in eggs.
    \item In a separate bowl, combine flour, baking soda, and salt.
    \item Add flour mixture bit-by-bit to other bowl. Mix well.
    \item Mix in chocolate chips.
    \item Place ping-pong-sized balls of dough onto cookie sheet, leaving space for them to spread out.
    \item Bake for 1.4 femto-Hubble times. Use the distance ladder result for softer cookies \cite[e.g.,][]{jwstSh0es} and CMB result \cite[e.g.][]{planck2018} for crisper cookies. For an in-depth discussion of cookie softness vs. crispiness, see \citet{cookieSoftness}.
\end{enumerate}

\end{multicols}

\vspace*{-3em}

\section*{Acknowledgments}
This recipe has been adapted from the NIST Metric Kitchen's  Chocolate Chip Cookies recipe \citep{nistCookies}. 
\textbf{Please note that the authors have not tested this recipe for evolving dark energy. Proceed with caution.}

\vfill
\newpage

\begin{multicols}{2}

%----------------------------------------------------------------------------------------
%	REFERENCE LIST
%----------------------------------------------------------------------------------------

\bibliographystyle{apalike}
\bibliography{main}

@misc{nistCookies, 
    author = {{NIST Metric Kitchen}},
    title = {Chocolate Chip Cookies},
    year={2010},
    note={\url{https://www.nist.gov/pml/owm/metric-si/metric-kitchen/metric-kitchen-recipe-gallery/metric-kitchen-chocolate-chip-cookies}}
}

@book{Bolton_1900, 
    place={Easton, PA}, 
    title={Evolution of the Thermometer}, 
    publisher={The Chemistry Publishing Co.}, 
    author={Bolton, Henry Carrington}, 
    year={1900}
}

@ARTICLE{planck2018,
       author = {{Planck Collaboration} and {Aghanim}, N. and {Akrami}, Y. and {Ashdown}, M. and {Aumont}, J. and {Baccigalupi}, C. and {Ballardini}, M. and {Banday}, A.~J. and {Barreiro}, R.~B. and {Bartolo}, N. and {Basak}, S. and {Battye}, R. and {Benabed}, K. and {Bernard}, J.-P. and {Bersanelli}, M. and {Bielewicz}, P. and {Bock}, J.~J. and {Bond}, J.~R. and {Borrill}, J. and {Bouchet}, F.~R. and {Boulanger}, F. and {Bucher}, M. and {Burigana}, C. and {Butler}, R.~C. and {Calabrese}, E. and {Cardoso}, J.-F. and {Carron}, J. and {Challinor}, A. and {Chiang}, H.~C. and {Chluba}, J. and {Colombo}, L.~P.~L. and {Combet}, C. and {Contreras}, D. and {Crill}, B.~P. and {Cuttaia}, F. and {de Bernardis}, P. and {de Zotti}, G. and {Delabrouille}, J. and {Delouis}, J.-M. and {Di Valentino}, E. and {Diego}, J.~M. and {Dor{\'e}}, O. and {Douspis}, M. and {Ducout}, A. and {Dupac}, X. and {Dusini}, S. and {Efstathiou}, G. and {Elsner}, F. and {En{\ss}lin}, T.~A. and {Eriksen}, H.~K. and {Fantaye}, Y. and {Farhang}, M. and {Fergusson}, J. and {Fernandez-Cobos}, R. and {Finelli}, F. and {Forastieri}, F. and {Frailis}, M. and {Fraisse}, A.~A. and {Franceschi}, E. and {Frolov}, A. and {Galeotta}, S. and {Galli}, S. and {Ganga}, K. and {G{\'e}nova-Santos}, R.~T. and {Gerbino}, M. and {Ghosh}, T. and {Gonz{\'a}lez-Nuevo}, J. and {G{\'o}rski}, K.~M. and {Gratton}, S. and {Gruppuso}, A. and {Gudmundsson}, J.~E. and {Hamann}, J. and {Handley}, W. and {Hansen}, F.~K. and {Herranz}, D. and {Hildebrandt}, S.~R. and {Hivon}, E. and {Huang}, Z. and {Jaffe}, A.~H. and {Jones}, W.~C. and {Karakci}, A. and {Keih{\"a}nen}, E. and {Keskitalo}, R. and {Kiiveri}, K. and {Kim}, J. and {Kisner}, T.~S. and {Knox}, L. and {Krachmalnicoff}, N. and {Kunz}, M. and {Kurki-Suonio}, H. and {Lagache}, G. and {Lamarre}, J.-M. and {Lasenby}, A. and {Lattanzi}, M. and {Lawrence}, C.~R. and {Le Jeune}, M. and {Lemos}, P. and {Lesgourgues}, J. and {Levrier}, F. and {Lewis}, A. and {Liguori}, M. and {Lilje}, P.~B. and {Lilley}, M. and {Lindholm}, V. and {L{\'o}pez-Caniego}, M. and {Lubin}, P.~M. and {Ma}, Y.-Z. and {Mac{\'\i}as-P{\'e}rez}, J.~F. and {Maggio}, G. and {Maino}, D. and {Mandolesi}, N. and {Mangilli}, A. and {Marcos-Caballero}, A. and {Maris}, M. and {Martin}, P.~G. and {Martinelli}, M. and {Mart{\'\i}nez-Gonz{\'a}lez}, E. and {Matarrese}, S. and {Mauri}, N. and {McEwen}, J.~D. and {Meinhold}, P.~R. and {Melchiorri}, A. and {Mennella}, A. and {Migliaccio}, M. and {Millea}, M. and {Mitra}, S. and {Miville-Desch{\^e}nes}, M.-A. and {Molinari}, D. and {Montier}, L. and {Morgante}, G. and {Moss}, A. and {Natoli}, P. and {N{\o}rgaard-Nielsen}, H.~U. and {Pagano}, L. and {Paoletti}, D. and {Partridge}, B. and {Patanchon}, G. and {Peiris}, H.~V. and {Perrotta}, F. and {Pettorino}, V. and {Piacentini}, F. and {Polastri}, L. and {Polenta}, G. and {Puget}, J.-L. and {Rachen}, J.~P. and {Reinecke}, M. and {Remazeilles}, M. and {Renzi}, A. and {Rocha}, G. and {Rosset}, C. and {Roudier}, G. and {Rubi{\~n}o-Mart{\'\i}n}, J.~A. and {Ruiz-Granados}, B. and {Salvati}, L. and {Sandri}, M. and {Savelainen}, M. and {Scott}, D. and {Shellard}, E.~P.~S. and {Sirignano}, C. and {Sirri}, G. and {Spencer}, L.~D. and {Sunyaev}, R. and {Suur-Uski}, A.-S. and {Tauber}, J.~A. and {Tavagnacco}, D. and {Tenti}, M. and {Toffolatti}, L. and {Tomasi}, M. and {Trombetti}, T. and {Valenziano}, L. and {Valiviita}, J. and {Van Tent}, B. and {Vibert}, L. and {Vielva}, P. and {Villa}, F. and {Vittorio}, N. and {Wandelt}, B.~D. and {Wehus}, I.~K. and {White}, M. and {White}, S.~D.~M. and {Zacchei}, A. and {Zonca}, A.},
        title = "{Planck 2018 results. VI. Cosmological parameters}",
      journal = {\aap},
         year = 2020,
        month = sep,
       volume = {641},
          eid = {A6},
        pages = {A6},
          doi = {10.1051/0004-6361/201833910},
archivePrefix = {arXiv},
       eprint = {1807.06209},
 primaryClass = {astro-ph.CO},
       adsurl = {https://ui.adsabs.harvard.edu/abs/2020A&A...641A...6P}
}

@ARTICLE{cookieSoftness,
       author = {{Rosu}, Sophie},
        title = "{All about Cookies: The perfect compromise between softness and crispiness}",
      journal = {arXiv e-prints},
         year = 2025,
        month = mar,
          eid = {arXiv:2503.23114},
        pages = {arXiv:2503.23114},
          doi = {10.48550/arXiv.2503.23114},
archivePrefix = {arXiv},
       eprint = {2503.23114},
 primaryClass = {astro-ph.SR},
       adsurl = {https://ui.adsabs.harvard.edu/abs/2025arXiv250323114R}
}

%----------------------------------------------------------------------------------------

\end{multicols}

\end{document}